\documentclass[preprint,12pt]{elsarticle}

\usepackage{amssymb}
\usepackage{amsmath}
\usepackage{algorithm}
\usepackage{algpseudocode}
\usepackage{algorithmicx}
\usepackage{algpseudocode}
\algdef{SE}[DOWHILE]{Do}{doWhile}{\algorithmicdo}[1]{\algorithmicwhile\ #1}%

\usepackage{color}

\journal{}

\begin{document}

\begin{frontmatter}



\title{Discovering Block Structure in Networks}


\author[inst1]{Rudy Arthur}

\affiliation[inst1]{organization={University of Exeter, Department of Computer Science},
            addressline={Stocker Rd}, 
            city={Exeter},
            postcode={EX4 4PY}, 
            country={UK}
}

\begin{abstract}
A generalization of modularity, called block modularity, is defined. This is a quality function which evaluates a label assignment against an arbitrary block pattern. 
Therefore, unlike standard modularity or its variants, arbitrary network structures can be compared and an optimal block matrix can be determined. Some simple algorithms for optimising 
block modularity are described and applied on networks with planted structure. In many cases the planted structure is recovered. Cases where it is not are analysed and it is found that 
strong degree-correlations explain the planted structure so that the discovered pattern is more `surprising' than the planted one under the configuration model. Some well studied networks 
are analysed with this new method, which is found to automatically deconstruct the network in a very useful way for creating a summary of its key features.
\end{abstract}




\end{frontmatter}


\section{Introduction}
\label{sec:intro}

Modularity \cite{newman2006modularity} is a function which takes a label assignment on the nodes of a network and returns a score evaluating how effectively the label assignment partitions the network into non-overlapping communities. Given a network with (weighted) adjacency matrix $A_{ij}$ and the labelling function $c(i)$ mapping nodes, $i$, to community labels, modularity is defined as
\begin{equation}\label{eqn:modularity}
Q_{\text{Newman}} = \frac{1}{2E} \sum_{ij} \left( A_{ij} - \gamma \frac{k_i k_j
}{2E} \right) \delta( c(i), c(j) ).
\end{equation}
$\sum_{ij} A_{ij} = 2E$ is the total number of edges, $k_i$ is the degree of the node $i$ and $\gamma$ is the so-called resolution parameter \cite{lambiotte2008laplacian}, set to $1$ throughout this work. The sum over $A_{ij}$ measures the fraction of within community edges in the observed network and the sum over degrees gives the fraction of within community edges expected under a degree preserving randomization of the network, known as the configuration model \cite{newman2003structure}. Despite some well-known issues identifying small communities (the so-called resolution limit \cite{fortunato2007resolution}) modularity maximization is the basis for a number of very popular community detection algorithms e.g. \cite{blondel2008fast, traag2019louvain}.

Modularity has been extended in a number of ways to identify communities in structured networks. 
Various authors \cite{guimera2004modularity, barber2007modularity, murata2009detecting} have given a definition of modularity appropriate for community detection in 
bipartite networks and 
\cite{arthur2020modularity} gives a definition of modularity appropriate for community detection in the unipartite projection of bipartite networks. 
A definition of modularity appropriate for 
finding multi-core-periphery structure is given in \cite{kojaku2017finding}. 
\cite{chen2014anti} defines ``anti-modularity" for finding sets of unconnected nodes, see also 
\cite{li2021anomaly} which uses modularity in a similar way as \cite{chen2014anti} to find approximately bipartite node sets. 
The original paper defining modularity \cite{newman2006modularity} has, at time of writing, over 11000 citations, 
so clearly modularity is an important and well used tool for community detection in networks.
\begin{figure}
    \centering
    \includegraphics[width=0.8\textwidth]{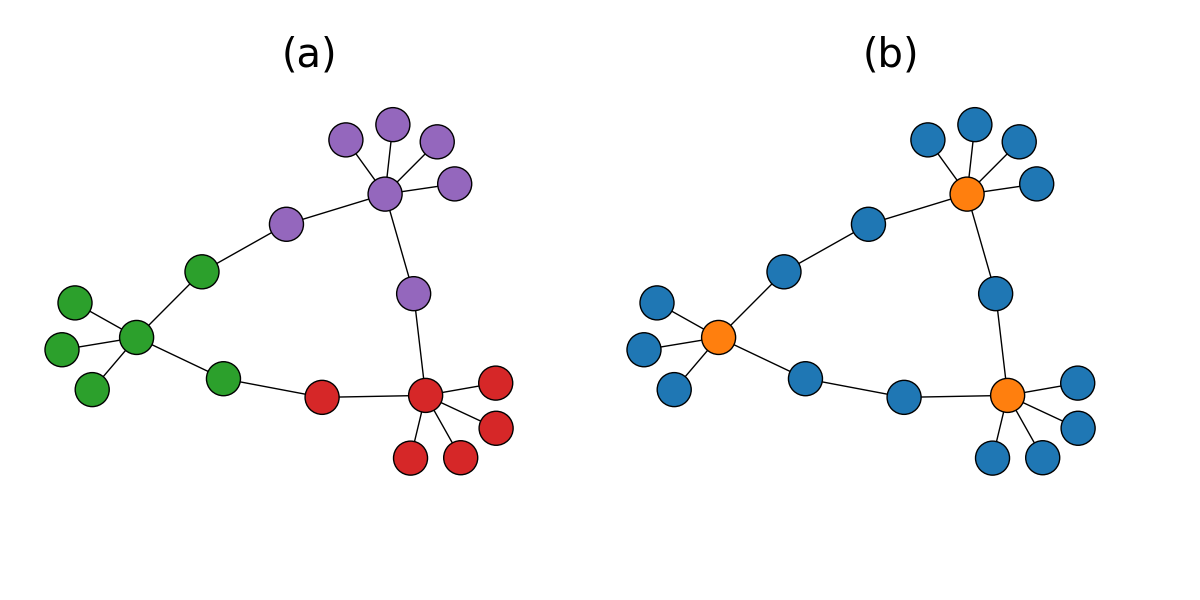}
    \caption{Same network with two different node labellings, indicated by colours.  (a) emphasises community structure and (b) bipartite structure. }
    \label{fig:fig1}
\end{figure}

Most work identifies a target structure, like non-overlapping communities or core and periphery sets, 
and aims to find a label assignment that maximises modularity or one of its  variants. However, consider Figure \ref{fig:fig1}. The same network is 
shown with two different labellings to emphasise either its approximate community structure or its approximate bipartite structure. 
While either of these labellings might be useful 
in different problem settings, this paper tries to answer the question of which one is `best' in the sense of `least expected under the configuration model'. 

Section \ref{sec:modmat} generalises modularity to arbitrary network structures which are specified by a block matrix $B$. Section \ref{sec:algo} describes algorithms to optimise this generalised modularity at fixed $B$ and then to optimise $B$ itself. These algorithms are applied to synthethic and real networks in Sections \ref{sec:algo} and \ref{sec:real}. Section \ref{sec:conc} summarises the results and suggests some directions for future work.

\section{Block Modularity}\label{sec:modmat}

Consider a network with $N$ nodes labelled into $N_B$ groups or `blocks'. Define the \textbf{matrix of modularity} $Q$ as the $N_B \times N_B$ matrix with elements
\begin{equation}\label{eqn:modularitymtx}
Q_{ab} = \sum_{ij} \left( A_{ij} - \gamma \frac{k_i k_j
}{2E} \right) \delta( c(i), a ) \delta( c(j), b )
\end{equation}
The standard modularity, equation \ref{eqn:modularity}, is equal to the trace of this matrix divided by $2E$. Following \cite{blondel2008fast}, it is convenient to change the sum over nodes in equation \ref{eqn:modularitymtx} to a sum over blocks. Defining 
\begin{align}
   \Sigma_{ab} &= \sum_{i \in a, j \in b} A_{ij} \\
   T_{a} &= \sum_{i \in a} k_i
\end{align}
lets us re-write equation \ref{eqn:modularitymtx} as
\begin{equation}
    Q_{ab} =  \Sigma_{ab} - \gamma \frac{T_a T_b}{2E}
\end{equation}
When there are more connections between $a$ and $b$ than the configuration model would predict $Q_{ab} > 0$ and when there are fewer $Q_{ab} < 0$. Thus, the sign and magnitude of $Q_{ab}$ is a measure of how `surprising' the edge density between node sets $a$ and $b$ is, relative to the configuration model. Large positive values correspond to an unexpected excess and large negative values to a deficit.

Define \textbf{block modularity} as
\begin{equation}\label{eqn:bm}
    Q(B) = \frac{1}{2E} \sum_{ab} Q_{ab} B_{ab}
\end{equation}
Here $B$ is a $N_B \times N_B$ matrix with entries equal to $\pm 1$. To gain some intuition it is helpful to consider the block matrices
\begin{equation*}
    B_0 = \begin{pmatrix}
    1 & -1 \\
    -1 & 1 \\
    \end{pmatrix} \text{ and }
    B_1 = \begin{pmatrix}
    -1 & 1 \\
    1 & -1 \\
    \end{pmatrix}
\end{equation*}
With nodes split into two blocks, with labels 0 and 1, $Q(B_0)$ will be large when there is an excess of edges within node sets and a deficit of edges between them - 
this is the usual non-overlapping two community structure. $Q(B_1)$ is the opposite, large when there is an excess of edges between node sets and a deficit within them. 
Therefore $Q(B_1)$ will be large for networks which are bipartite or approximately so. 

Other modularity formulations can be recovered by taking different values for the block matrix $B$. The standard equation \ref{eqn:modularity}, can be recovered with
$B_{ab} = \delta_{ab}$ for example. Other formulations can be recovered by substituting the corresponding block pattern. If we split the label $a$ into a parity label
$x_a$ and community label $c_a$ then a modularity definition suitable for a bipartite graph made of multiple communities, after \cite{barber2007modularity}, can be obtained using
$B_{x_a,c_a;x_b,c_b} = \delta_{c_a c_b}(1 - \delta_{x_ax_b})$. Similarly, the multi-core-periphery modularity of \cite{kojaku2017finding} can be recovered with
$B_{x_a,c_a;x_b,c_b} = \delta_{c_a c_b}(x_a+x_b-x_ax_b)$. In contrast to most other definitions of modularity, we will use $B$ matrices with values $\{+1, -1\}$ rather than $\{1,0\}$.
This is to give equal weight to excesses and deficits of connections between blocks.

\begin{figure}
    \centering
    \includegraphics[width=0.5\textwidth]{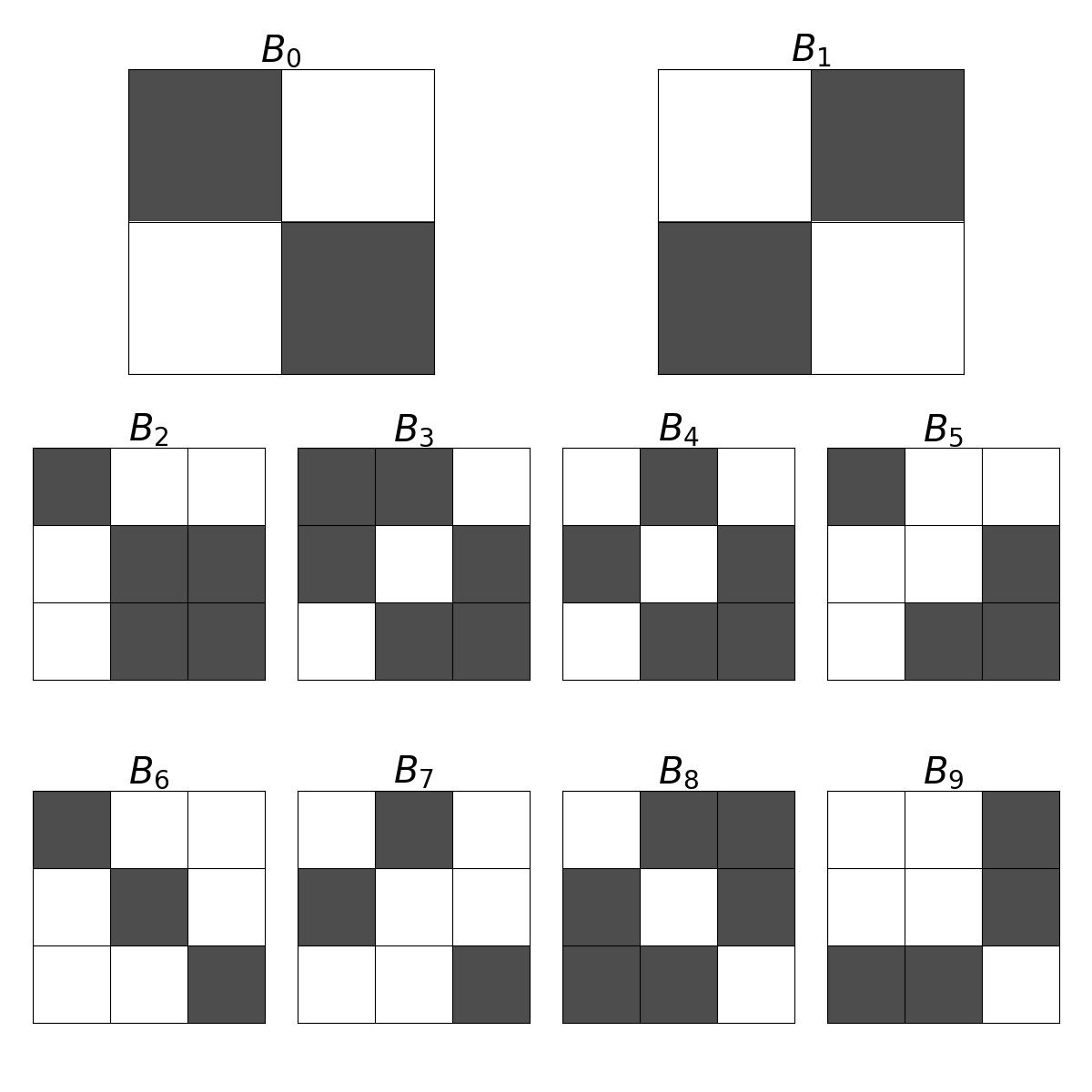}
    \caption{Allowed $2 \times 2$ and $3 \times 3$ block patterns, after \cite{kojaku2018core}.}
    \label{fig:blocks}
\end{figure}
The work of \cite{kojaku2018core} is the most similar to the above. They provide limits on the types of block structure the configuration model can detect. By 
enforcing symmetry and the identity
\begin{equation}\label{eqn:consistency}
    \sum_a \Sigma_{ab} = \sum_a \frac{T_a T_b}{2E}
\end{equation}
certain types of block matrix are forbidden. For example, it is not possible to simultaneously have an excess within and between all node sets. Defining black cells as ones 
where $Q_{ab} > 0$ and white cells where $Q_{ab} < 0$, equation \ref{eqn:consistency} is equivalent to forbidding completely white or black rows or columns so, in particular, 
a $2 \times 2$ core-periphery block pattern is forbidden. The allowed $2 \times 2$ and $3 \times 3$ patterns are shown in figure \ref{fig:blocks}.

The maximum value of standard modularity is 1 \cite{brandes2007modularity}. To understand the maximum value of $Q(B)$, let us follow \cite{danon2005comparing} and consider a block matrix $B$ with $+1$ on the diagonal and $-1$ elsewhere for a network of $N_B$ cliques with the canonical labelling. The diagonal terms contribute
\[
 \frac{1}{N_B}  \left(1 - \frac{1}{N_B} \right)
\]
and the off-diagonals give
\[
 -\frac{1}{N_B^2} 
\]
So that
\begin{equation}
Q(B)_{max} = \frac{N_B}{N_B}\left(1 - \frac{1}{N_B} \right) + \frac{N_B(N_B-1)}{N_B^2} = 2 - \frac{2}{N_B}
\end{equation}
which converges to $2$ for large $N_B$. The value is $2$ instead of the standard modularity bound of $1$ due to counting the deficits as well as the excesses. 
Like standard modularity, the upper bound is achieved only in the limit of very large $N_B$ thus, like standard modularity, $Q(B)$ will tend to be larger for higher $N_B$. 
In this work we compare $Q(B)$ at fixed $N_B$, comparing values at different $N_B$ should be done cautiously.

For any fixed $B$, optimising equation \ref{eqn:bm} will find the label assignment on the nodes that best matches that structure. For example
$B_6$ is the block pattern of 3 isolated communities. $Q(B_6)$ for the network in Figure \ref{fig:fig1} will be large given labels shown in 1(a). 
$B_1$ is the block pattern of a bipartite network.
$Q(B_1)$ for the same network will be large for the label assignment shown in 1(b). Given there are a finite number of allowed block patterns 
we can find the optimal label assignment for every $B$ and compare all of the maximised values of $Q(B)$ for different $B$. The block matrix that gives the maximum 
$Q(B)$ score is the least expected under the configuration model and therefore represents the structure in the network which can be least well explained by degree correlation. 
In the following sections we will give some examples that show how this optimal structure matrix can be useful in characterising networks.

\section{Algorithms for Finding Block Patterns}\label{sec:algo}

\begin{algorithm}
\caption{$\text{Label Swap}(B, T_0=0.01, k_{max}=1000)$ }\label{alg:cap}
\begin{algorithmic}
\State $\text{moves} \gets 1$
\State $k \gets -1$
\State Let $\vec{n}$ be the list of nodes
\State Let $c(n)$ be the label of node $n$ 
\While{$\text{moves} > 0$ or $k < k_{max}$}
    \State $\text{moves} \gets 0$
    \State $k \gets k+1$
    \State $T = T_0\left( \frac{k_{max} - k}{k_{max}} \right)^2$
    \State Randomly shuffle the list of nodes $\vec{n} \gets \vec{n}'$
    \For{$n$ in $\vec{n}'$} 
        \For{  $a$ in block labels where $a\neq c(n)$ }
        \State Compute $dQ(a)$, the change in $Q(B)$ when $c(n) \gets a$
        \EndFor
        \State $dq \gets \text{max}\left( dQ(a) \right)$
        \State $c_{max} \gets \text{argmax}\left( dQ(a) \right)$
        \If{$dq > 0$} 
            \State $\text{moves} \gets \text{moves} + 1$
            \State $c(n) \gets c_{max}$
        \ElsIf { $k < k_{max}$ and $r < \exp(dq/T)$ }
            \State $c(n) \gets c_{max}$
        \EndIf
    \EndFor
\EndWhile
\State \textbf{return} $Q(B)$
\end{algorithmic}
\end{algorithm}

For fixed $B$ and some initial labelling of the nodes $c(i)$, the optimisation algorithm \ref{alg:cap} finds a labelling with high $Q(B)$. 
The algorithm performs simulated annealing with quadratic cooling, swapping node labels to increase $Q(B)$, where moves that decrease $Q(B)$ are allowed at higher temperature. 
$r$ is a random number and typical parameters are $T_0 = 0.1$, $k_{max} = 100$. The algorithm \ref{alg:cap} performs better than a greedy method (setting $T = 0$, $k_{max} = 0$) 
in most cases. 

\begin{figure}
    \centering
    \includegraphics[width=\textwidth]{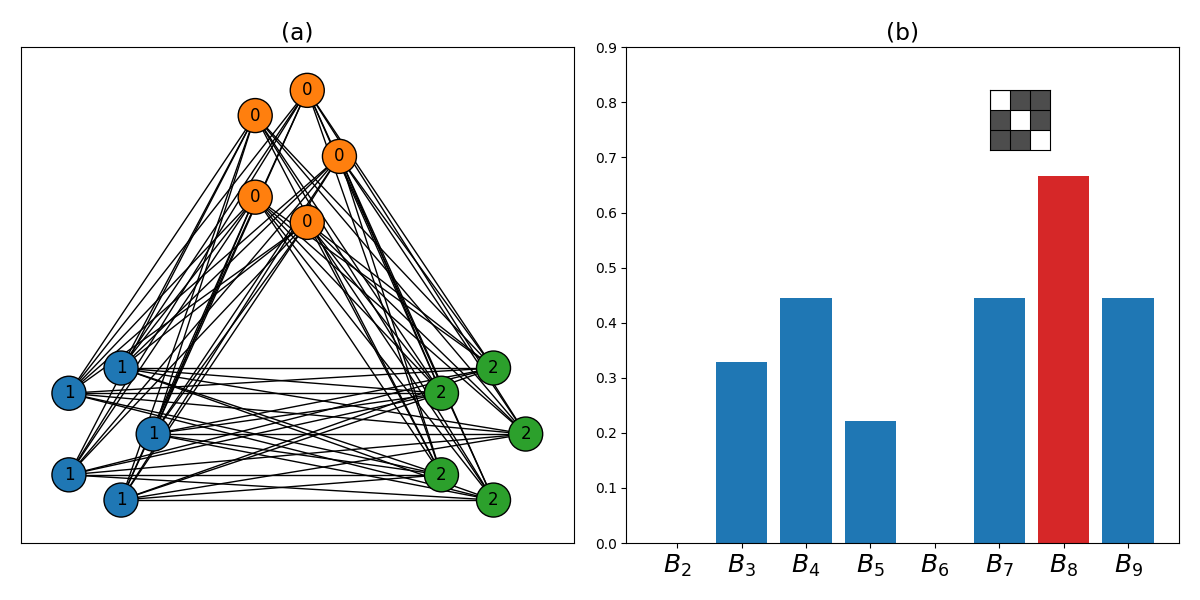}
    \caption{(a) A tripartite network with the optimal labelling under $Q(B_8)$, (b) the values of $Q(B_i)$ for all $3 \times 3$ block patterns, the optimum is achieved at $B_8$. }
    \label{fig:tripartite}
\end{figure}
Figure \ref{fig:tripartite} (a) shows a tripartite network. Running algorithm \ref{alg:cap} using each of the allowed $3 \times 3$ block patterns in Figure \ref{fig:blocks} 
gives the optimal values of $Q(B_i)$ shown in Figure \ref{fig:tripartite} (b). The maximum $Q(B)$ is achieved for $B = B_8$, which is the block pattern corresponding to the tripartite structure of the network, with the optimal labelling for $Q(B_8)$ indicated by the colours in Figure \ref{fig:tripartite} (a). 

\begin{figure}
    \centering
    \includegraphics[width=\textwidth]{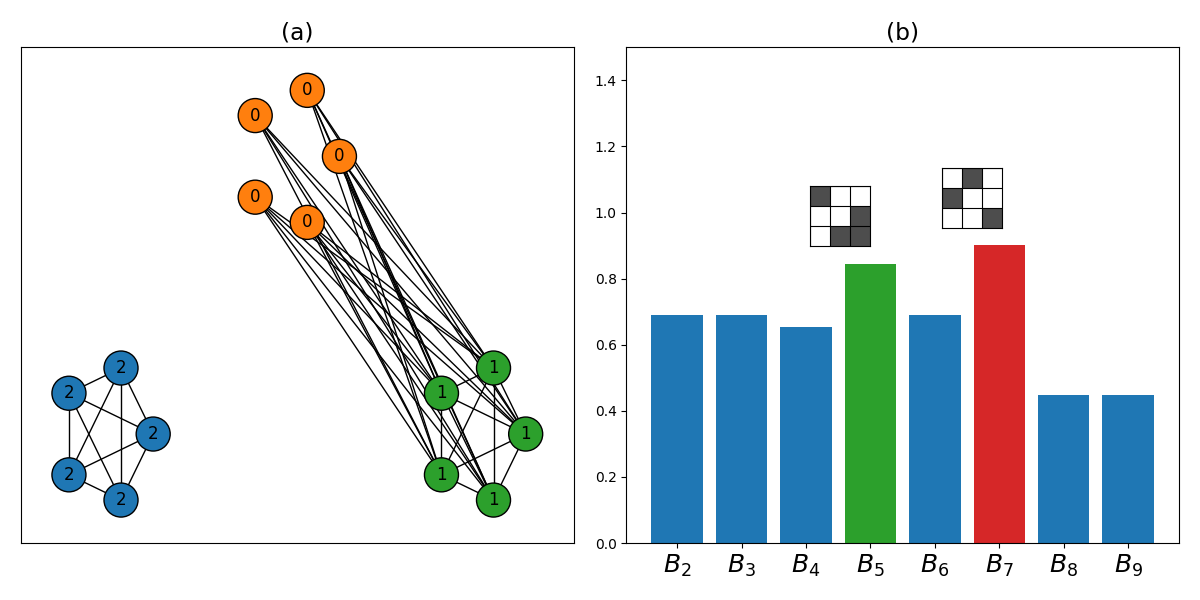}
    \caption{(a) A core-periphery network (pattern $B_5$) with the optimal labelling under $Q(B_7)$, 
    (b) the values of $Q(B_i)$ for all $3 \times 3$ block patterns. }
    \label{fig:ccp}
\end{figure}
It is not always the case that the block pattern `planted' in the network is the one recovered by optimising $Q(B)$. Figure \ref{fig:ccp} (a) shows a network consisting a
core-periphery with an isolated communtiy, which corresponds to block pattern $B_5$. The same process is applied and the optimal $Q(B)$ is achieved with $B = B_7$ rather than $B_5$. 
This is further analysed in the next section and \ref{sec:appendixA} where it is shown that, due to the high degrees of the core nodes, connections between core nodes 
are expected under a degree preserving 
randomisation. This means that patterns with bipartite structure (as in $B_7$) are favoured since they admit a labelling that is `more surprising'.

\begin{figure}
    \centering
    \makebox[\textwidth][c]{\includegraphics[width=1.2\textwidth]{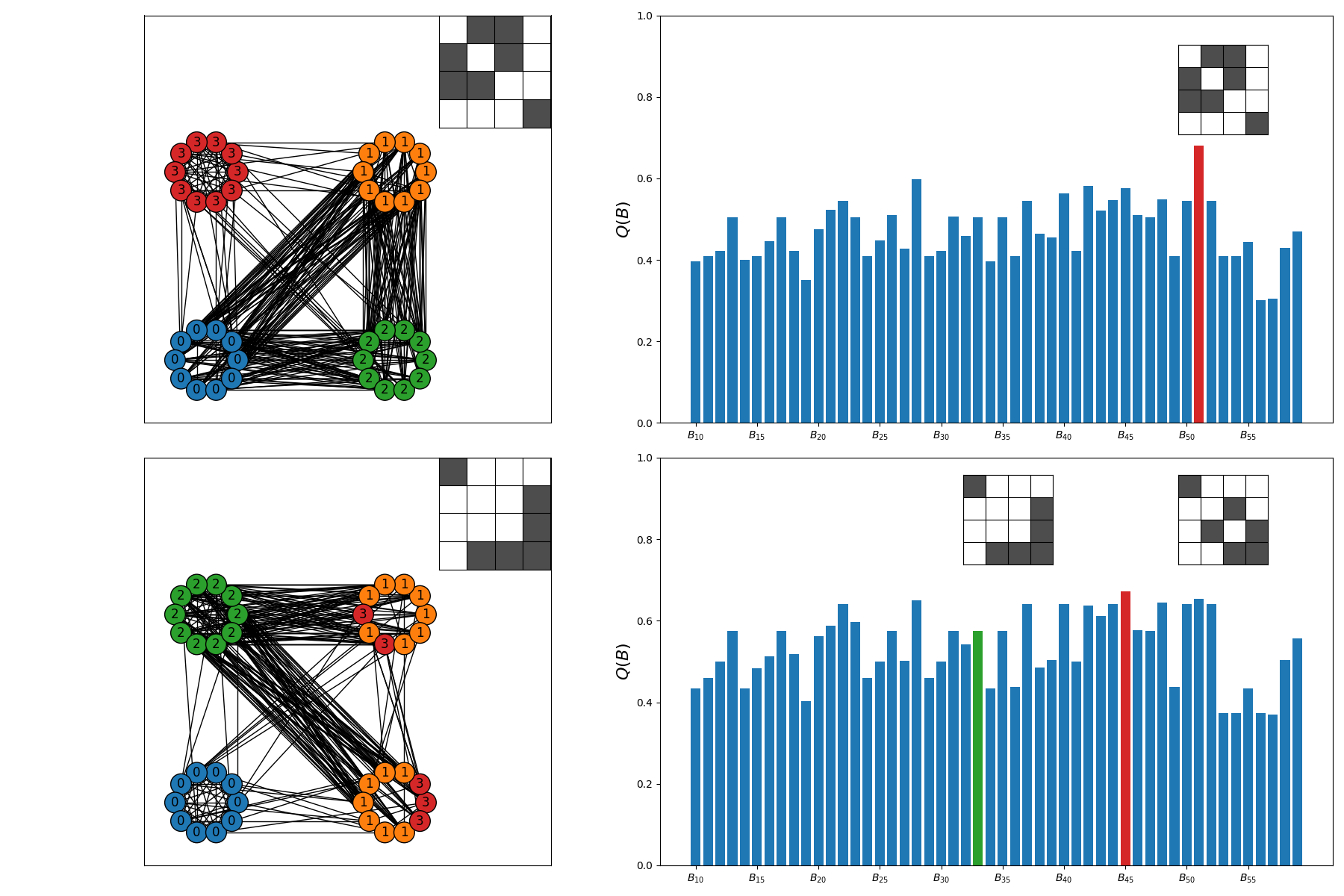}}
    \caption{Left: The network, the block pattern used in its construction and the optimal labelling under the optimal $B$ identified on the right. Right: $Q(B)$ for every allowed $4 \times 4$ block pattern. }
    \label{fig:fourbyfour}
\end{figure}
To generate more complicated networks with planted block structures, for $N_B$ blocks, $n$ nodes per block, construct an $N_B \times N_B$ density matrix $P$, where $P_{ab}$ is the 
probability of a link between nodes in block $a$ and block $b$. Figure \ref{fig:fourbyfour} shows two networks where $N_B = 4$, $n = 10$ and $P$ is constructed by replacing black 
elements with $0.8$ and white elements with $0.1$ in the corresponding block pattern. This is just the Stochastic Block Model (SBM), see e.g. the review of \cite{lee2019review}. 
In this work the SBM is only used to generate networks with interesting block structure, but the SBM has a close relationship with modularity maximisation. It was shown in 
\cite{newman2016community} that maximising the modularity is equivalent to finding the maximum likelihood estimate of the parameters of a particular type of SBM called the planted 
partition model. The approach taken here of optimising equation \ref{eqn:bm} likely has some relationship with maximum likelihood estimation of the parameters of some SBM, 
we will discuss connections to the SBM further in Section \ref{sec:conc}.

The top row of Figure \ref{fig:fourbyfour} shows a network which consists of an isolated, dense community, loosely connected to a tripartite network.
 Running Algorithm \ref{alg:cap} for all possible $4 \times 4$ block patterns to find the optimal labelling for each gives the result on the right. 
 The planted structure is recovered as the structure corresponding to the maximum $Q(B)$.

It is appropriate at this stage to look at the performance of Algorithm \ref{alg:cap}. This algorithm, at any value of $T_0$, can get stuck at local
maxima, much like other modularity maximisation algorithms \cite{blondel2008fast}. Slower cooling schedules (larger $k_{max}$) typically find better maxima. For 
the network shown in the bottom of Figure \ref{fig:fourbyfour} optimising $Q(B)$ with the block partition shown in the figure on the left, 
using the greedy algorithm ($T_0 = 0$) we find $31/100$ runs achieve the optimal label assignment. Using the annealing algorithm ($T_0 = 0.01$, $k_{max} = 100$) this goes up 
to $49/100$. These numbers are broadly representative of other networks and
block patterns, with annealing finding the optimum a factor of $\sim 2-5$ times more often in most cases. 
Thus it is recommended (and implemented in Figure \ref{fig:fourbyfour} and elsewhere) to run Algorithm \ref{alg:cap} a number of times, $N_r$, and choose the run with the largest 
value of $Q(B)$. $N_r = 20$ is found to be sufficient in the cases considered in this work. 

Again, it is not always the case that the planted structure is recovered. 
The bottom row of Figure \ref{fig:fourbyfour} shows a network constructed as an isolated community, loosely connected to a core-periphery network, 
where the core has two distinct peripheries. The optimal $B$ and label assignments are not the canonical ones implied by the block matrix used to construct the network, 
a number of other block patterns admit labellings with higher $Q(B)$. The largest $Q(B)$ is found with the pattern shown the on right of Figure \ref{fig:fourbyfour}.
The nearly isolated community is recovered exactly but instead of the `core double periphery' pattern there is a very small 
core connected to one periphery which itself forms one half of a nearly bipartite pair.  

To show explicitly how and why core-periphery structure can
`vanish' under block modularity, consider a fully connected clique of $M$ nodes, all sharing the label $0$,
connected to a periphery of $qM$ nodes, labelled $1$. To satisfy the consistency condition, equation \ref{eqn:consistency},
also add a disconnected clique of $bM$ nodes. In \ref{sec:appendixA} it is shown that
$Q(B_5)$, the optimal block modularity for the planted structure is greater than $Q(B_7)$ \textbf{only if} $b > q^2$.

The reqirement $b > q^2$ means that if the periphery is large (high $q$)
or if the core-periphery makes up the majority of the network (low $b$) then the core-periphery block pattern can be a sub-optimal description of the network structure under
the configuration model. 
Ultimately, this condition derives from squaring the degree sum of the core nodes. Since these nodes have very high degree,
under a degree preserving randomisation it is not unlikely that they are connected to each other. 
In the network since only a subset of these `core' edges are intra-block connections
the high connectedness in the core is expected and doesn't contribute to the `surprise' measured by $Q(B)$. From this we can conclude that, as well as the conditions given by 
equation \ref{eqn:consistency}, for core-periphery structure to exist (under the configuration model) requires either a relatively small periphery or that 
the core-periphery only forms a relatively small part of the 
overall network. The result is that core-periphery structure, even if explicitly planted, can give lower $Q(B)$ than using a block pattern where the core is missing. 
The result is analogous to \cite{kojaku2018core}, certain block patterns may intuitively appear to be optimal, but under the configuration model they are not, due to the importance of 
degree correlations. 

\begin{algorithm}
\caption{Simulated Annealing($N_B, T_0 = 0.01, k_{max}=100$)}\label{alg:sim}
\begin{algorithmic}
\State $B = I_{N_B \times N_B}$
\State $Q = \text{Label Swap}(B)$
\While{$T > 0$}
    \State $T = T_0\left( \frac{k_{max} - k}{k_{max}} \right)^2$
    \Do
    \State For a random element of $B$
    \State $B_{ij} \gets -B_{ij}$
    \If{$i \neq j$} \State $B_{ji} \gets -B_{ji}$ \EndIf
  \doWhile{equation \ref{eqn:consistency} is false} 
    \State $Q_n = \text{Label Swap}(B)$
    \If{$Q_n > Q$ or $r < \exp( (Q_n-Q)/T )$}
        \State $Q = Q_n$
    \Else
        \State Undo flip
    \EndIf
\EndWhile
\end{algorithmic}
\end{algorithm}

\begin{figure}
    \centering
    \makebox[\textwidth][c]{\includegraphics[width=\textwidth]{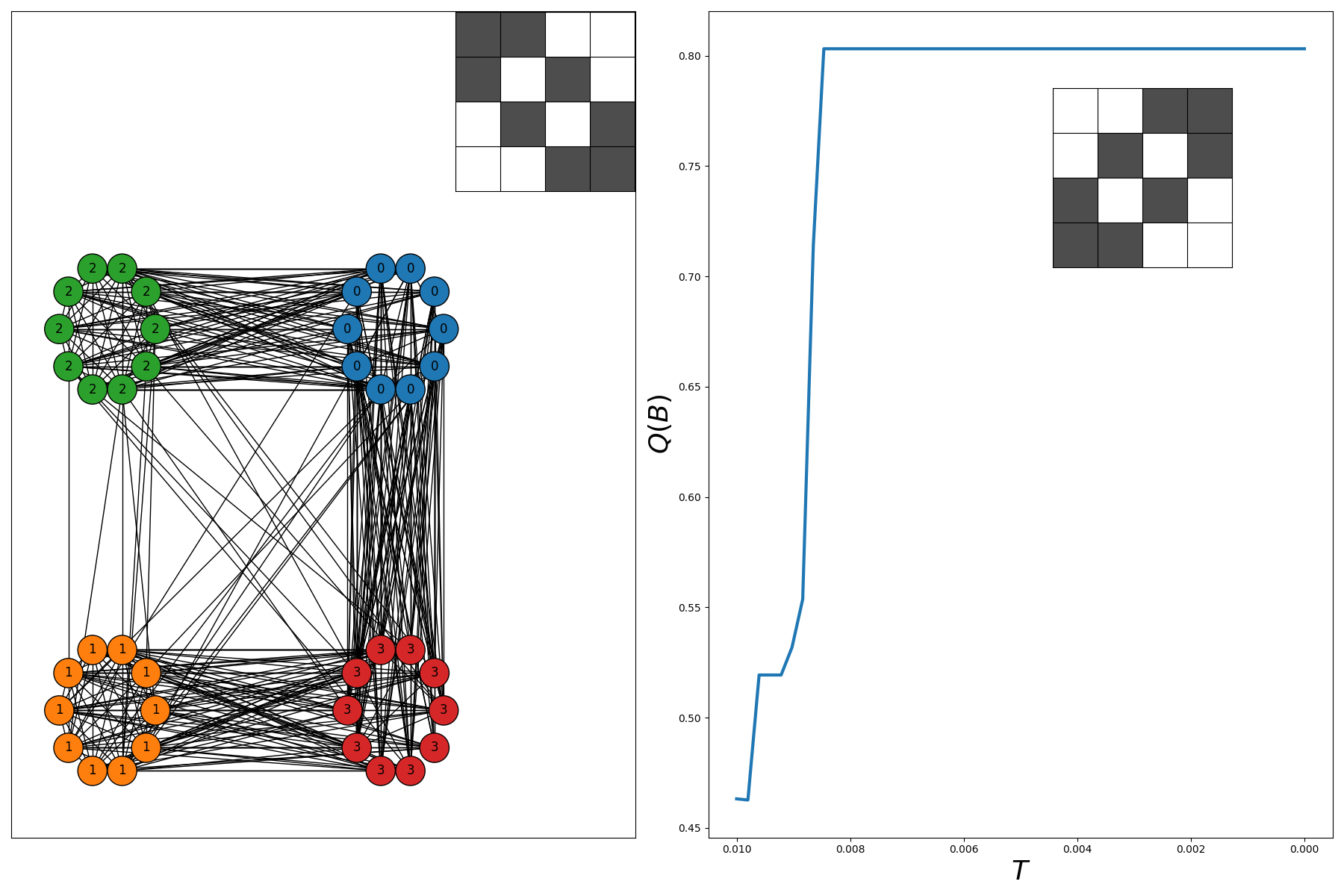}}
    \caption{Left: The network and the block pattern used in its construction (which is also the optimal block pattern identified by Algorithm \ref{alg:sim}). 
    The network is labelled according to to the block pattern found by Algorithm \ref{alg:sim}, shown on the right. 
    Right: Shows the block pattern found by Algorithm \ref{alg:cap} and $Q(B)$ as a function of temperature. }
    \label{fig:anneal}
\end{figure}

The number of allowed block patterns grows quite rapidly with $N_B$ and an exhaustive search becomes infeasible. Inspired by the similarity of these block matrices to spin systems, 
Algorithm \ref{alg:sim} is a simulated annealing approach to finding the optimal block pattern. Figure \ref{fig:anneal} shows the final block pattern and label set found by this algorithm 
for $T_{init} = 0.01$ and $1000$ steps of quadratic cooling. In practice the final results of algorithms \ref{alg:cap} and \ref{alg:sim} are fairly insensitive to the exact cooling scheme, 
starting temperature and number of cooling steps. Figure \ref{fig:anneal} shows that algorithm \ref{alg:sim} recovers an equivalent block pattern to the one used to generate the network, 
where a permutation of the labels turns the pattern on the right into the one on the left. The optimal labelling for this block pattern $B$ is the one expected 
based on the planted structure. Algorithm \ref{alg:sim}, like algorithm \ref{alg:cap}, can become
stuck at local maxima. The same solution - repeated, independent runs - is used to alleviate this problem.
Algorithms \ref{alg:sim} and \ref{alg:cap} will find optimal label assignments and block patterns but are somewhat inefficient. In this work the intention is to understand $Q(B)$ rather 
than find the best possible algorithm to optimise it, so algorithmic improvements are left for future work.

\section{Real Networks}\label{sec:real}

\begin{figure}
    \centering
    \includegraphics[width=\textwidth]{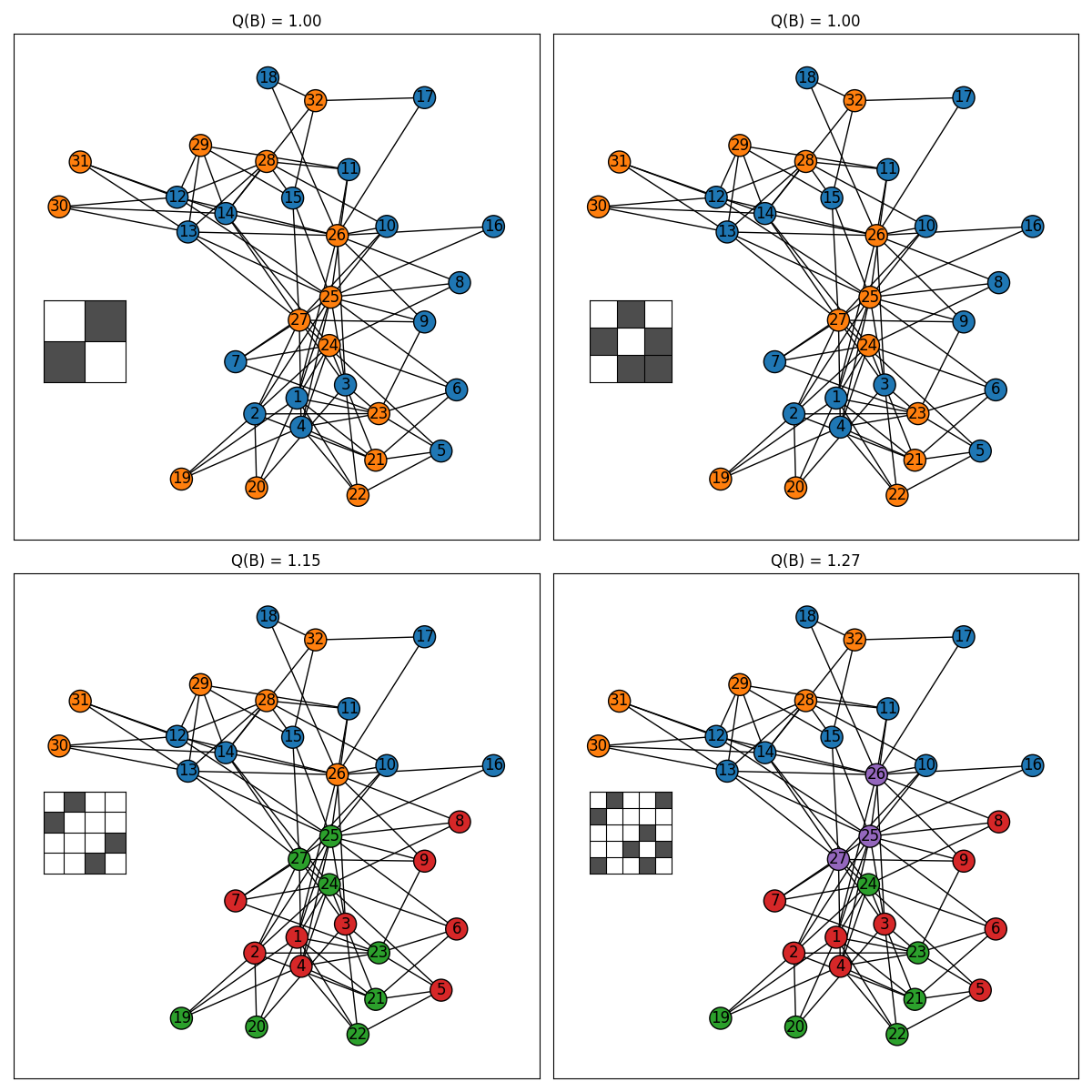}
    \caption{The optimal block pattern, labelling and value of $Q(B)$ for the `Southern Women' network for $N_B  = \{2,3,4,5\}$. The map of colours to block indices is \{0:blue, 1:orange, 2:green, 3:red, 4:purple, 5:brown\} }
    \label{fig:south}
\end{figure}

In this section algorithm \ref{alg:sim} will be run on some well known empirical networks obtained from the KONECT database \cite{kunegis2013konect}. Since the optimal number of blocks is unknown, the algorithm is run for a range of $N_B$. The first network to be analysed is the Southern Women network \cite{davis2009deep}, a bipartite network consisting of 18 women (labelled 1 through 18) and 14 events (labelled 19 through 32). Typical analysis of this network \cite{everett2013dual} identifies 2 or more communities of women and 2 or more classes of event. The block optimisation is summarised in Figure \ref{fig:south}. For $N_B=2$ the algorithm identifies the (exact) bipartite structure of the network.  $N_B=3$ demonstrates some interesting behaviour; the optimal labelling does not include any labels for the third block. A block pattern with lower $N_B$ can outperform a higher $N_B$ pattern. Algorithms \ref{alg:cap} and \ref{alg:sim} discover this fact by returning an optimal partition that only uses a subset of the allowed node labels.

$N_B = 4$ splits the events into two groups (red and orange) and splits the women into two groups (red and blue) with green events predominantly attended by red women and blue women attending orange events. $N_B = 5$ refines this picture, with 3 classes of event and 2 communities of women. Blue women attend orange and purple events, while red women attend green and purple events. This is also shown by the block matrix. The purple events bridge the two separate bipartite communities. This network has been analysed in great detail, e.g. in \cite{everett2013dual}, and it is remarkable that this kind of structure can be detected fairly automatically by optimising $Q(B)$.

\begin{figure}
    \centering
    \includegraphics[height=0.9\textheight]{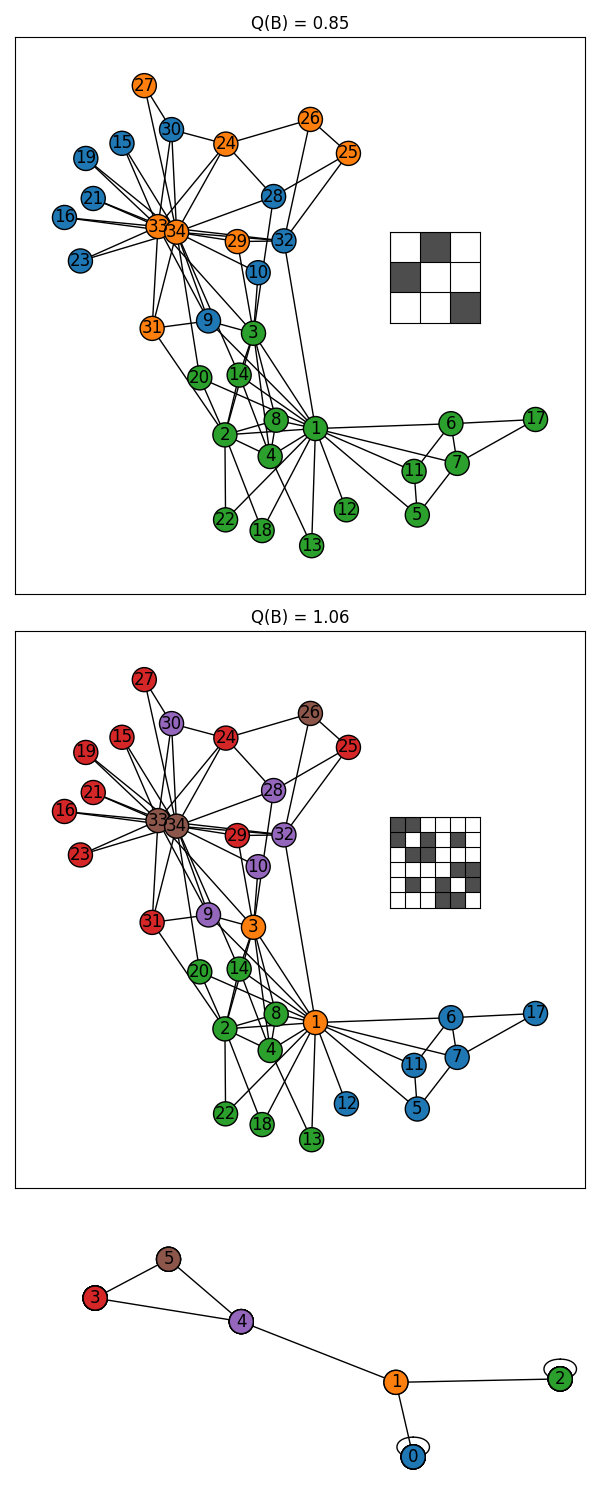}
    \caption{The optimal block pattern, labelling and value of $Q(B)$ for the Karate Club network with $N_B  = \{3,6\}$. The map of colours to block indices is \{0:blue, 1:orange, 2:green, 3:red, 4:purple, 5:brown\}. The bottom shows a representation of the $6 \times 6$ block matrix as a network. }
    \label{fig:karate}
\end{figure}

Another commonly studied network is Zachary's karate club \cite{zachary1977information}, a social network split into two communities due to a dispute between the club's instructor (node 33) and administrator (node 1). Using $N_B=2$ recovers the usual two community pattern, Figure \ref{fig:karate} shows results for $N_B = \{3,6\}$. Higher $Q(B)$ can be obtained for higher $N_B$ but there is a balance to strike between optimisation and interpretability of the block matrix. It is also the case that optimisation of larger block patterns is slower and more prone to get stuck in local minima.

The $3 \times 3$ pattern is a community centered around the administrator and a bipartite community of the other members, where the blue nodes have very few ties with anyone other than the instructor. The optimal $6 \times 6$ pattern is shown in the middle panel. For large $N_B$ it can be hard to interpret the block matrix from the binary pattern alone, so this pattern is visualised as a network at the bottom of the figure, where a clearer picture emerges. Blocks 2 and 0 are two independent communities which only interact with the administrator's faction (block 1). The instructor's faction, block 5, interacts with a `loyalist' block (3) and another block (4) which retains some ties to the administrator. Important to note is the fact that unless a block has a self loop it should not be considered a `community'. For example, the nodes in block 3 have very few connections to each other, members of this group are only connected to each other only via the instructor's faction.

\begin{figure}
    \centering
    \includegraphics[height=\textwidth]{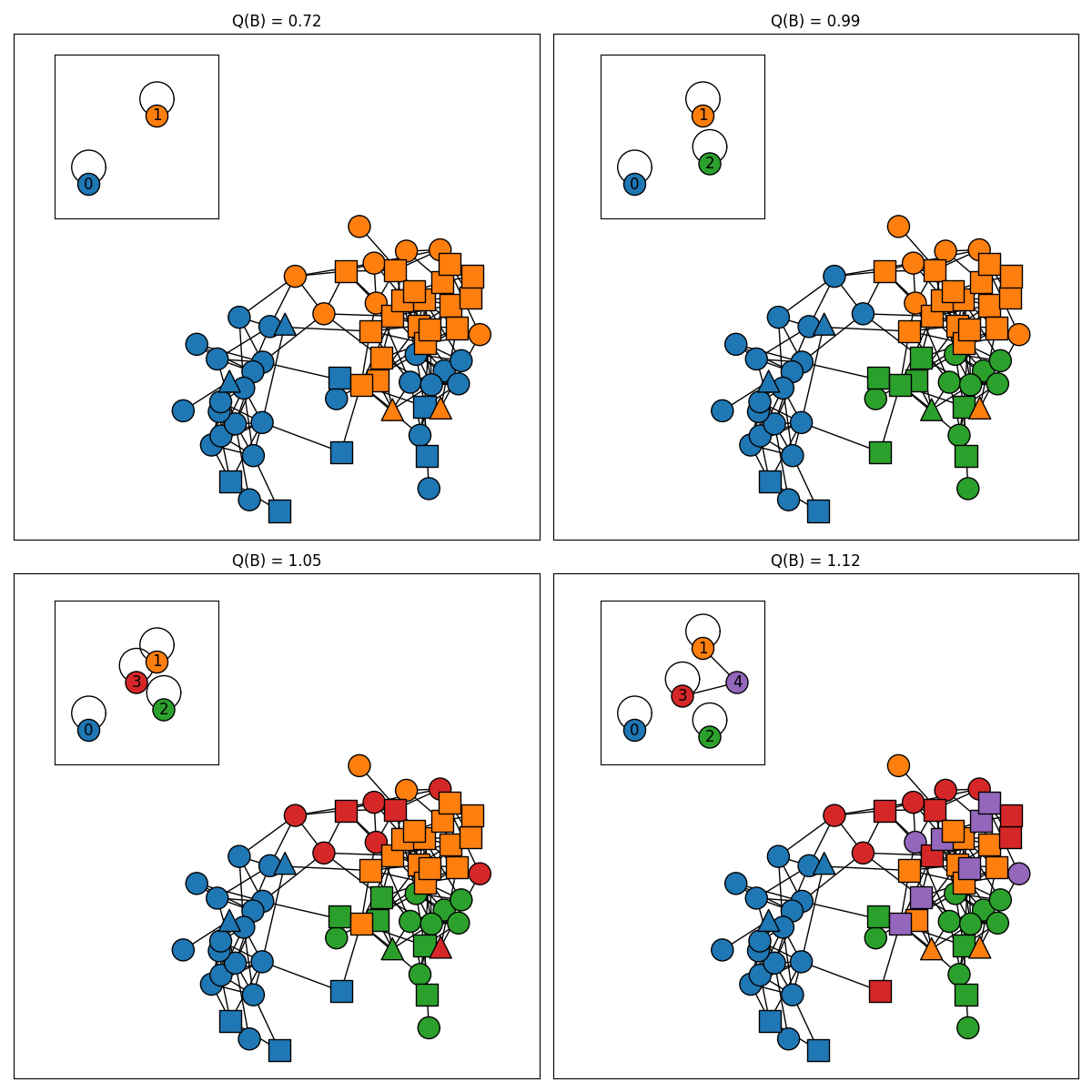}
    \caption{The optimal block pattern, labelling and value of $Q(B)$ for the Dolphin network with $N_B  = \{2,3,4,5\}$. The map of colours to block indices is \{0:blue, 1:orange, 2:green, 3:red, 4:purple\}. Squares are female dolphins, circles are male and triangles are unknown. The inset figure shows the network diagram corresponding to the optimal block pattern, see Figure \ref{fig:karate}. }
    \label{fig:dolphin}
\end{figure}
The final example is Figure \ref{fig:dolphin} which shows the dolphin social network from \cite{lusseau2003bottlenose}, which is suggested to be an example of a multiple core 
and periphery structure in \cite{kojaku2018core}. The optimal block patterns for $N_B \in \{2,3,4\}$ are $N_B$ isolated communities with the labelling shown. For $N_B=5$ the 
pattern is a little more interesting, the all female group 1 when $N_B = 4$ splits in two and loses some members to the mixed red group. Blocks 1 and 3 now form two cores 
connected to a shared periphery, block 4. Unlike Figure \ref{fig:ccp} where a planted core-core-periphery structure could not be detected, as discussed in \ref{sec:algo}, 
if a core-periphery pattern forms a small enough subset of the total network, as here, it can be detected. This kind of structure is also seen in Figure \ref{fig:karate}, where 
block 1 is the shared periphery of 0 and 2.

\section{Conclusions}\label{sec:conc}

This paper presents a generalization of modularity, called block modularity, which provides a framework for detecting arbitrary structure in networks. Because different 
structures are evaluated with the same quality function, these structures can be compared and an optimal one identified, meaning fewer assumptions need to be made about 
the network. Searching for particular patterns, like community structure or core-periphery, is still a useful thing to do, however block modularity and an algorithm like 
\ref{alg:sim} can approach the network `blind' and discover potentially interesting and unexpected structures. 

Section \ref{sec:real} shows that optimal block patterns can help create a `narrative' account of a complex network. They can also be counter-intuitive, in particular it seems that 
core-periphery structure is sometimes elusive. The configuration model is a fairly powerful null model and degree correlations `explain' a lot of network structure. It would be 
interesting future research to explore in general what circumstances core-periphery networks may be better explained by other block 
patterns, under the configuration model. 

To apply this method on very large networks (web pages, social networks), faster optimisation algorithms are required. Algorithm \ref{alg:cap} requires $O(N^2)$ operations per 
temperature increment and both algorithms can become trapped in local minima. Direct enumeration of all block patterns is preferable where possible, though seems to only be practical 
for $N_B < 5$. Ideally, like the Louvain and Leiden methods \cite{blondel2008fast, traag2019louvain}, algorithms could be developed which are not only fast, but which do not 
require the number of blocks $N_B$ to be specified in advance, allowing a completely blind approach. 

This approach shares many of the problems of standard modularity maximisation. For example the resolution limit identified by \cite{fortunato2007resolution}
as preventing the detection of small communities. There is also the fact that `modular' partitions can be found even for random networks \cite{guimera2004modularity}. As emphasised in 
\cite{peel2022statistical} and \cite{peixoto2021descriptive}, modularity maximisation is not an inferential approach, like maximum likelihood estimation of a SBM, but is 
descriptive. The label assignment, and here also the structure matrix $B$, found by modularity maximisation answers the question of which sets of nodes, in a specific network, 
have more or fewer inter-connections than a degree preserving randomisation would predict. When applied to any network, even a random one,
the maximum modularity score may be low but \textit{something} will be found, since we are asking for a description of that network.

The aim of this paper is not to add to the zoo of community detection methods. \cite{ghasemian2019evaluating} has performed a thorough
study of numerous different community detection algorithms and their relative performance. \cite{young2018universality} has shown that the very general SBM framework
subsumes many different `mesoscopic pattern extraction' problems, including community detection by modularity maximisation. Following \cite{newman2016community}
or \cite{young2018universality} it is likely that the maximisation of equation (6) has some relationship with the SBM. Understanding these connections
and comparing the outputs of the methods described here (or some refinement of them) against modern SBM techniques would be interesting future work. This paper
proposes a way to unify the many different modularity-like functions in the literature; demonstrates some unexpected and interesting consequences for multi-core-periphery networks
and shows how block matrices provide a nice `summary' of a network. Recent critism of modularity based clustering \cite{peel2022statistical, peixoto2021descriptive} raises
many interesting points about the drawbacks of the method and the misunderstanding of its results in applications. However, while modularity remains a popular approach
to structure detection I hope that the unifying framework described here will help researchers and practitioners to better understand these methods and use them appropriately.

\appendix

\section{Core-periphery}\label{sec:appendixA}

Consider a fully connected clique of $M$ nodes, all sharing the label $0$,
connected to a periphery, labelled $1$, of $qM$ nodes and a disconnected clique of $bM$ nodes.
\begin{equation*}
\Sigma = \begin{pmatrix}
M^2 & qM^2 & 0\\
qM^2 & 0 & 0\\
0 & 0 & bM^2
\end{pmatrix}
\end{equation*}
\begin{equation*}
T = \begin{pmatrix}
(1+q)M^2 & qM^2 & bM^2
\end{pmatrix}
\end{equation*}
\begin{equation*}
2E =M^2((1+q) + (b+q))
\end{equation*}
Gives
\begin{equation*}
Q = M^2\left( 
\begin{pmatrix}
1 & q & 0\\
q & 0 & 0\\
0 & 0 & b
\end{pmatrix}
-
\begin{pmatrix}
(1+q)^2 & q(1+q) & b(1+q)\\
q(1+q) & q^2 & bq\\
b(1+q) & qp & b^2
\end{pmatrix} \frac{1}{(b+q) + (1+q)}
\right)
\end{equation*}

The matrix element $Q_{11}$
\begin{equation*}
M^2 \left(  -\frac{q^2}{(1+q) + (b+q)}\right)
\end{equation*}
is always negative. The matrix elements $Q_{01}, Q_{10}$ 
\begin{equation*}
M^2 \left( q - \frac{q(1+q)}{(1+q) + (b+q)}\right)
\end{equation*}
are always positive, as long as
\begin{equation*}
(b+q) > 0
\end{equation*}
which is always true, since $b$ and $q$ are positive and non-zero for non-trivial networks.

The interesting term is $Q_{00}$ which is
\begin{equation*}
M^2 \left( 1 - \frac{(1+q)^2}{(1+q) + (b+q)}\right)
\end{equation*}
This is positive if
\begin{equation*}
b > q^2 
\end{equation*}
and otherwise negative. If $Q_{00}$ is negative then a bipartite block pattern with 
\begin{equation*}
\begin{pmatrix}
B_{00} & B_{01}\\
B_{10} & B_{11}
\end{pmatrix} = 
\begin{pmatrix}
-1 & 1\\
1 & -1
\end{pmatrix}
\end{equation*}
will give higher $Q(B)$ than the `core-periphery' block pattern.

 \bibliographystyle{elsarticle-num} 
 \bibliography{manuscript}





\end{document}